# Deep Learning Enhanced Extended Depth-of-Field for Thick Blood-Film Malaria High-Throughput Microscopy

Petru Manescu[1], Lydia Neary- Zajiczek[1], Michael J. Shaw[1], Muna Elmi [1], Remy Claveau [1], Vijay Pawar [1], John Shawe-Taylor [1], Iasonas Kokkinos [1], Mandayam A. Srinivasan [1], Ikeoluwa Lagunju [2,3], Olugbemiro Sodeinde [2,3], Biobele J. Brown [2,3], Delmiro Fernandez-Reyes [1,2,3]

[1] Department of Computer Science, Faculty of Engineering, University College London, London, United Kingdom
[2] Department of Paediatrics, College of Medicine University of Ibadan, University College Hospital, Ibadan, Nigeria
[3] Childhood Malaria Research Group, College of Medicine University of Ibadan, University College Hospital, Ibadan, Nigeria

**Abstract.** Fast accurate diagnosis of malaria is still a global health challenge for which automated digital-pathology approaches could provide scalable solutions amenable to be deployed in low-to-middle income countries. Here we address the problem of Extended Depth-of-Field (EDoF) in thick blood film microscopy for rapid automated malaria diagnosis. High magnification oil-objectives (100x) with large numerical aperture are usually preferred to resolve the fine structural details that help separate true parasites from distractors. However, such objectives have a very limited depth-of-field requiring the acquisition of a series of images at different focal planes per field of view (FOV). Current EDoF techniques based on multi-scale decompositions are time consuming and therefore not suited for high-throughput analysis of specimens. To overcome this challenge, we developed a new deep learning method based on Convolutional Neural Networks (EDoF-CNN) that is able to rapidly perform the extended depth-of-field while also enhancing the spatial resolution of the resulting fused image. We evaluated our approach using simulated low-resolution z-stacks from Giemsa-stained thick blood smears from patients presenting with Plasmodium falciparum malaria. The EDoF-CNN allows speed-up of our digital-pathology acquisition platform and significantly improves the quality of the EDoF compared to the traditional multi-scaled approaches when applied to lower resolution stacks corresponding to acquisitions with fewer focal planes, large camera pixel binning or lower magnification objectives (larger FOV). We use the parasite detection accuracy of a deep learning model on the EDoFs as a concrete, task-specific measure of performance of this approach.

# 1    Introduction

*Plasmodium falciparum* malaria remains one of the greatest world health burdens with over 200 million cases globally leading to half-million deaths annually, mostly among children [1]. Fast and accurate diagnosis is required for administering the proper treatment. Currently, the gold standard for malaria diagnosis is the microscopic evaluation of thick blood smears by trained pathologists. Unfortunately, visual inspection is time-consuming and relies on the availability of trained personnel. Recently, computer vision techniques have attempted to automatically detect malaria parasites in digitized microscopy of thick blood smears [2, 3]. Due to the small size of the ring-stage parasites in circulation ($\sim$ 2 to 3µm) high magnification oil-objectives (100x) with a large numerical aperture (typically 1.4 NA) are usually preferred. Since these objectives have a limited field of view (FOV) and a limited depth of field (DOF), multiple images at different focal planes (z-stack) are acquired. Ten to fourteen planes are needed to capture the entire thickness of the blood smear for one FOV. While some attempts try to use the entire stack for analysis [3], a resolved depth of field image is desired for both analysis purposes as well as visual inspection in the context of a computer aided system for malaria diagnosis. Despite their effectiveness, current microscopy fusion techniques which rely on multi-scale decompositions (MSD) [4] are computationally expensive and thus not suited for prompt malaria diagnosis of specimens, a task that can necessitate the analysis of hundreds of FOVs.

In this paper, we explore the use of Convolutional Neural Networks (CNN) to obtain fast resolved depth of field images. We investigate how this approach can, at the same time, enhance the digital spatial resolution of the fused in-focus image, enabling the automated diagnosis of malaria using cheaper, lower magnification objectives and thus larger FOVs. The next section presents existing works related to this study, followed by a detailed description of our approach. Section 4 presents and discusses the preliminary results while the final section concludes the paper.

# 2    Related work

Deep learning methods based on CNNs have been recently developed to achieve state-of-the-art multi-focus fusion of natural scene images [5, 6]. These approaches formulate the fusion of two partially blurred images of the same content as a two-class classification problem. They are based on Siamese-like CNN models trained to output the corresponding focus map used for fusion. It remains unclear how well these methods scale to fuse more than two partially out-of-focus images.

At the same time, CNNs have been used to increase digital spatial resolution in natural images [7] and more recently in the field of microscopy [8]. The spatial resolution enhancement is formulated as a regression problem: a low-resolution image is passed through a CNN model whose weights are optimized during training to minimize the pixel-wise difference between the CNN output and the corresponding high resolution target image. In this context, we have re-formulated the multi-focus fusion problem as a regression task to achieve the fusion of the different focal planes while enhancing the digital spatial resolution of the fused image at the same time.

## 3    Methods and materials

### 3.1    CNN-based Extended Depth of Field

We modified an existing encoder-decoder architecture [9] for the EDoF-CNN which consisted of three 2D convolutional layers (encoder), followed by nine residual layers and three 2D transposed convolutional layers (decoder) with a *tanh* activation function at the end. To accommodate multiple focal planes as an input to the CNN we propose the following two approaches:

**EDoF-CNN-3D.** In this method, the z-stack was considered as a 3D volume and consequently, the encoder part of the CNN architecture was modified by replacing the two-dimensional convolutions with three-dimensional ones. The output tensor was flattened on the z-axis before the residual layers of the network by average pooling. The rest of the architecture remained unchanged.

**EDoF-CNN-Max.** This approach combines the idea behind the Siamese networks and the one behind the wavelet-based MSD EDoF [4]. Each focal plane is passed through the encoder part of the network and the maximum of the activation values are selected before going through the residual layers. Figure 1 summarizes the two architectures.

### 3.2    Experimental set-up

From high resolution acquired stacks we simulated lower resolution ones corresponding to different faster acquisition scenarios. The EDoF-CNN models were then trained to fuse the simulated stacks while enhancing their spatial resolution (Fig. 2). We used an Olympus (BX63) microscope with a 100X/1.4NA objective and a color digital camera (PCO) to acquire multiple z-stacks of Giemsa-stained thick blood smears from our clinics. For each stack, 14 focal planes were acquired at different focal depths with a z-step of a 0.5 μm, ensuring, thus, a total depth of 7 μm for each FOV. The EDOF was computed for every stack using a wavelet decomposition method [4]. Specifically, each z-plane was decomposed using a 12 level "sym8" wavelet. For each level and sub-band the coefficients with the maximum values were chosen among the 14 decomposed focal planes. Following a sub-band consistency check, the inverse wavelet transform was applied to the selected coefficients. The obtained EDOF images were used as a target image (ground-truth) to train our EDoF-CNN models. All the images in this study were converted to grayscale. The following three low-resolution acquisition scenarios were simulated:

**A. Larger z-step.** We simulated larger z-steps (smaller acquisition time) during the acquisition by selecting only a part of the z-planes from the initial 14 plane z-stack: a z-step of 1.5 μm which corresponds to 5 z-planes (i.e. selecting only the 1st plane, the 4th one, the 7th, etc.) and a z-step of 2.5 μm corresponding to only 3 z-planes.

**B. Binning.** A common practice in digital microscopy is to group a number of camera pixels forming a "larger" pixel ensuring a lower noise level at the expense of the loss of spatial resolution. We simulated a 4x4 binning by down-sampling the selected stacks by a factor of 4 using local averaging.

**C. Lower magnification.** A lower magnification objective (40x/0.6 NA) acquisition was simulated: first, the entire z-stacks were convolved with a simulated three-dimensional point-spread-function (PSF) [10] corresponding to a 0.6 NA and then down-sampled by a factor of 2.5. For the three scenarios described above, the two EDoF-CNN models implemented in Tensorflow [11] were trained with z-stacks of 512x512 pixels (33.3μm x 33.3μm) per focal plane each. The mean squared pixel-wise difference was chosen as a loss function. The training dataset consisted of 3900 FOVs coming from 13 thick blood smears showing different levels of parasitemia.

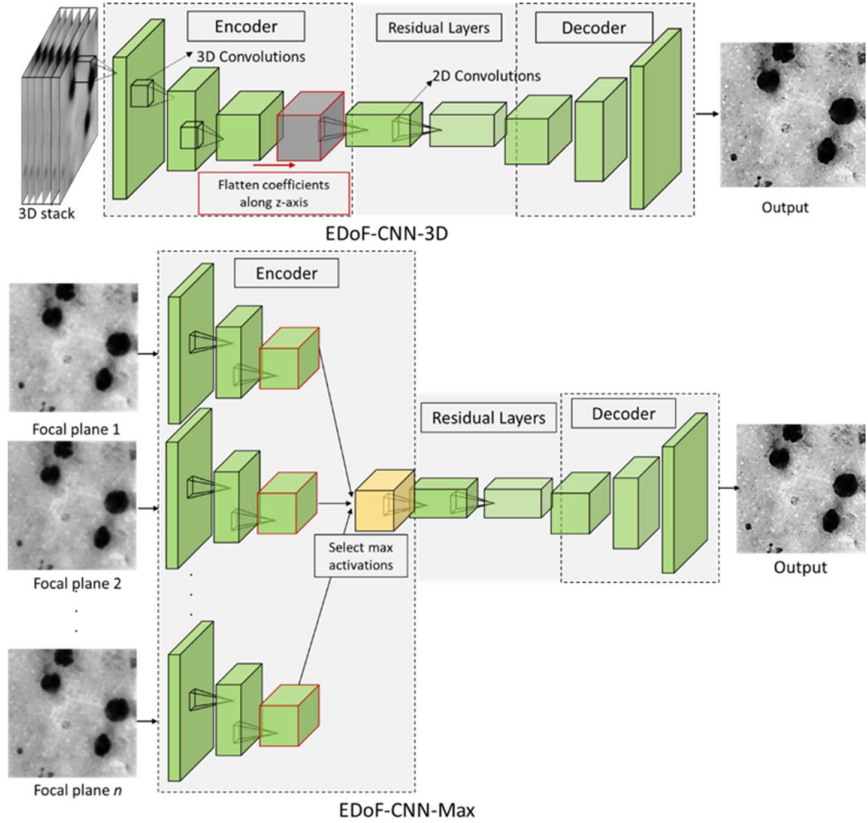

**Fig. 1.** Proposed CNN architectures for the EDoF-CNN approach.

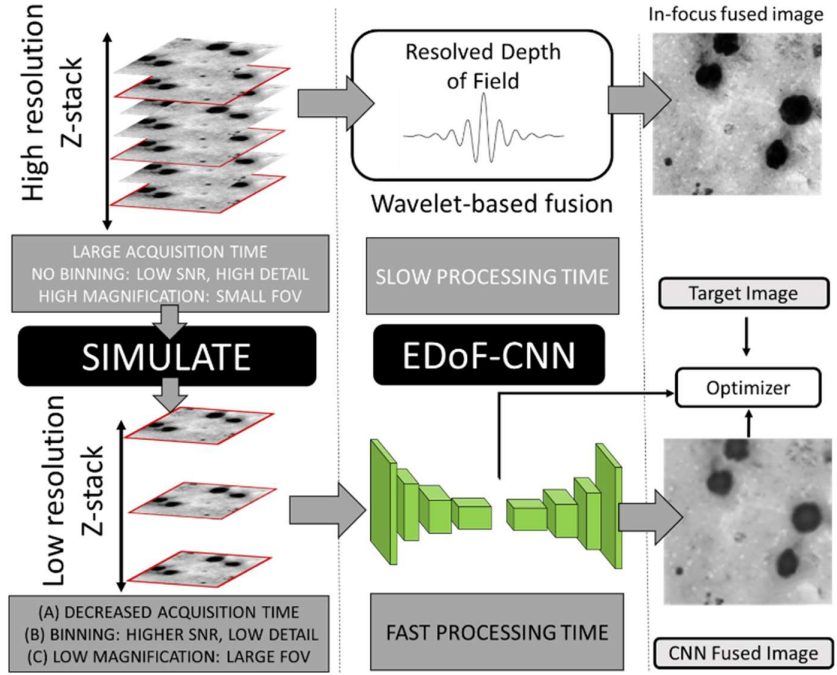

**Fig. 2.** EDoF-CNN experimental set-up.

### 3.3 Evaluation set-up

A test dataset of unseen 100 stacks of 512x512 pixels for each plane, acquired from a thick blood film of a patient with high parasitemia was used to evaluate the improvement of our EDoF-CNN methods with respect to the wavelet-based EDoF for larger z-step, binning and low magnification scenarios. We first compared the fused images with the target high resolution fused images in terms of image quality using the SSIM index [12]. We next implemented a simple parasite region segmentation such as in [2]. The grayscale fused images were first binarized using the Otsu method [13] and then binary blobs with areas larger than $3\mu m^2$ and smaller than $0.5\mu m^2$ were removed. This method was both applied to the target high resolution EDOF image as well as to the fused images for each scenario. The Dice index was used to compare the similarity of the segmentations [14]. Next, we used a CNN-based object detector [15] trained to find parasites in thick blood smears. The accuracy of the detector on the fused images was computed with respect to the parasites detected on the high-resolution EDoF target images. Finally, the computation times needed to fuse 100 stacks with 3 and 5 z-planes for the wavelet-based EDOF method (single process and parallelized) and the two EDoF-CNN methods (inference time) were measured.

## 4 Results

The EDoF-CNN methods improved the quality of the fused images compared to the wavelet-based EDOF method in terms of SSIM (from 0.77 to 0.81) (Table 1) and of the segmentation Dice scores (0.60 to 0.66) (Table 2). The SSIM is not as sensitive to blurring and loss of detail as the Dice index is. The improvement of the Dice scores becomes even more significant in the case of the *binning* (0.58 to 0.66) simulation and the *lower magnification* one (0.50 to 0.66) when applying the segmentation based on the Otsu thresholding (Table 2). Our EDoF-CNN models enhanced the spatial resolution of the fused images in the case of a simulated 40x/0.6NA z-stack almost matching the quality of the initial 100x/1.4NA magnification (Fig. 3). The parasite detection recall on images obtained from a simulated 40x magnification and 5 focal planes fused with the EDoF-CNN-3D method is boosted from 38% to 73%, reaching almost the same as the recall corresponding to a 100x magnification (Table 3).

**Table 1.** Average SSIM values with respect to the target high resolution EDOF images.

| Number of z-planes | Wavelet EDOF | EDoF-CNN- 3D | EDoF-CNN- Max |
|---|---|---|---|
| 3 | 0.74 ± 0.06 | 0.77±0.05 | 0.78±0.04 |
| 5 | 0.77 ± 0.03 | **0.80±0.02** | **0.81±0.02** |
| **4x4 Binning** | | | |
| 3 | 0.76 ± 0.03 | 0.76±0.04 | **0.77±0.03** |
| 5 | 0.78 ± 0.02 | 0.76±0.04 | **0.79±0.02** |
| **40x/0.6NA Magnification** | | | |
| 3 | 0.74 ± 0.03 | **0.76±0.03** | 0.76±0.03 |
| 5 | 0.74 ± 0.02 | 0.78±0.02 | **0.78±0.02** |

**Table 2.** Average Dice scores of the segmentation accuracy with respect to the segmentation of the high resolution EDOF images

| Number of z-planes | Wavelet EDOF | EDoF-CNN- 3D | EdoF-CNN-Max |
|---|---|---|---|
| 3 | 0.56±0.17 | 0.60±0.17 | **0.62±0.15** |
| 5 | 0.60±0.17 | **0.67±0.15** | 0.66±0.16 |
| **4x4 Binning** | | | |
| 3 | 0.55±0.17 | 0.53±0.21 | **0.61±0.16** |
| 5 | 0.58±0.16 | 0.64±0.16 | **0.66±0.15** |
| **40x/0.6NA Magnification** | | | |
| 3 | 0.48 ± 0.17 | **0.63±0.15** | 0.62±0.15 |
| 5 | 0.50 ± 0.17 | **0.66±0.16** | 0.60±0.17 |

**Table 3.** Malaria parasite detection accuracy w.r.t the detection on the high resolution images

| Number of z-planes | Wavelet EDOF (Recall/Precision) | EDoF-CNN- 3D (Recall/Precision) | EDoF-CNN- Max (Recall/Precision) |
|---|---|---|---|
| 3 | 0.60/0.81 | 0.61/0.75 | **0.67/0.75** |
| 5 | 0.70/0.80 | **0.74/0.79** | 0.72/0.78 |
| **4x4 Binning** | | | |
| 3 | 0.51/0.80 | 0.61/0.76 | **0.67/0.73** |
| 5 | 0.55/0.80 | 0.65/0.78 | **0.73/0.75** |
| **40x/0.6NA Magnification** | | | |
| 3 | 0.34/0.71 | 0.70/0.72 | **0.71/0.71** |
| 5 | 0.38/0.72 | **0.73/0.76** | 0.72/0.74 |

These results show that, with our EDoF-CNN methods, larger fields of views acquired with 40x objectives could be used for automatic malaria parasite detection. Figure 4 shows an example of such a detection output in the case of a 40x objective. The EDoF-CNN methods take advantage of the fast GPU implementation and reduced the computational time by up to a factor of 11 compared to the single process wavelet-based implementation (Table 4). The EDoF-CNN-3D method was slower than the EDoF-CNN-Max one due to the 3D convolution operations.

**Table 4.** Computational times to perform the fusion of 100 stacks of 512x512 pixels. The Wavelet EDOF was implemented in Matlab 2018a using CPU (parallel) processing while the CNN methods use Tensorflow's GPU implementation. Tested on an Intel Core i9 3.1 GHZ CPU with a NVIDIA GeForce RTX GPU with 12 Gb of memory.

| Number of z-planes | Wavelet EDOF | Wavelet EDOF Parallel | EDoF-CNN-3D | EDoF-CNN-Max |
|---|---|---|---|---|
| 3 | 142 s | 37 s | 29 s | **18 s** |
| 5 | 234 s | 64 s | 31 s | **20 s** |

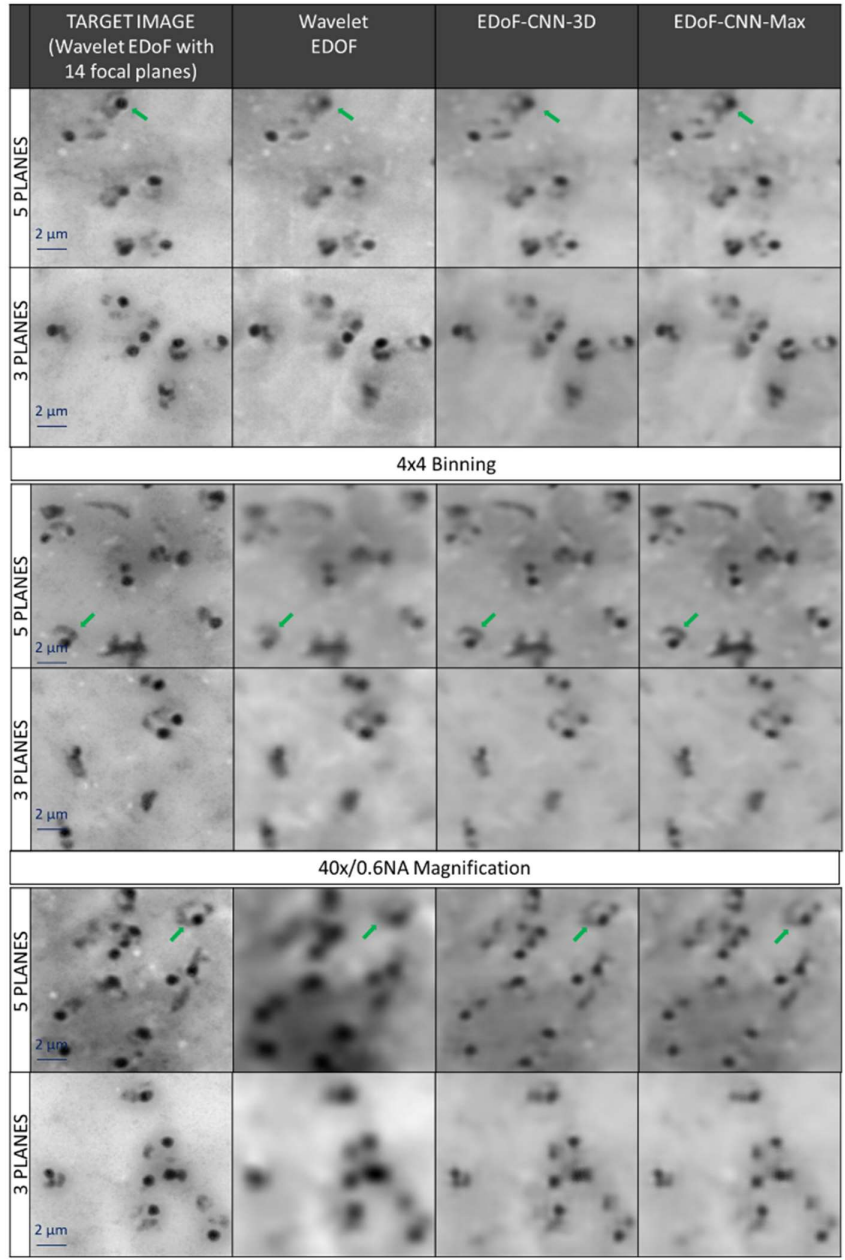

**Fig. 3.** Fused images using the different methods under the different scenarios (13µm x 13 µm). Green arrows indicate examples of fine details of the parasites recovered by the EDoF-CNN.

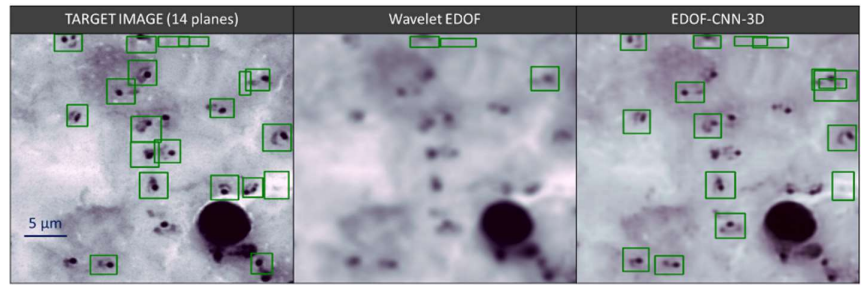

**Fig. 4.** Example of malaria parasite detection using a deep learning object detection method for the 40x/0.6NA magnification simulation. Grayscale fused images were re-colored.

## 5  Conclusion

We presented a new CNN framework able to rapidly compute extended depth of field images from z-stacks of thick blood films for automated malaria diagnosis. The proposed method has the benefit of enhancing the images spatial resolution at the same time. The current experiments were performed on grayscale stacks and simulated low resolution acquisitions. We plan on extending our approach to color images and apply it to improve the spatial resolution of acquired stacks at lower magnifications. The EDoF-CNN approach could thus allow the analysis of larger fields of views without losing important details, which is suitable for high-throughput microscopy.